# Title: Amorphous alloys surpass E/10 strength limit at extreme strain rates

**Authors:**

Wenqing Zhu[1*], Zhi Li[2*], Hua Shu[3], Huajian Gao[2, 4†] and Xiaoding Wei[1, 5, 6†]

**Affiliations:**

[1]State Key Laboratory for Turbulence and Complex System, Department of Mechanics and Engineering Science, College of Engineering, Peking University, Beijing 100871, China.

[2]School of Mechanical and Aerospace Engineering, College of Engineering, Nanyang Technological University, 70 Nanyang Drive, 637457, Singapore

[3]Shanghai Institute of Laser Plasma, China Academy of Engineering Physics, Shanghai, 201102, China.

[4]Institute of High Performance Computing, A*STAR, 138632, Singapore

[5]Beijing Innovation Center for Engineering Science and Advanced Technology, Peking University, Beijing 100871, China.

[6]Peking University Nanchang Innovation Institute, Nanchang 330000, China

*These authors contributed equally to this work.

†To whom correspondence and requests for materials should be addressed. E-mail: huajian.gao@ntu.edu.sg (H.G.); xdwei@pku.edu.cn (X.W.)

**Abstract:** Theoretical predictions of the ideal strength of materials range from $E/30$ to $E/10$ ($E$ is Young's modulus). However, despite intense interest over the last decade, the value of the ideal strength that can be attained experimentally for metals remains a mystery (*1-5*). In this study, we demonstrated the unprecedented strength of an amorphous Cu-Zr alloy that surpassed the $E/10$ limit. Laser-induced shock experiments were conducted on $Cu_{50}Zr_{50}$ to explore its strength and failure mechanisms at ultrahigh strain rates. The material demonstrated a high spall strength of 9.8 GPa, approximately 1/13 of its P-wave modulus (~ $E/6$), at strain rates greater than $10^7\,s^{-1}$, which sets a new record for the elastic limit of metallic materials. Electron microscopy and large-scale molecular dynamics simulations revealed that void nucleation and growth, not shear-banding, comprised the major failure mechanism for metallic glasses at extremely fast strain rates. A new model for void formation under the control of surface energy explained the rate dependence of the material strength. The results of this study reveal new possible ways to use the amorphous phase in nanostructured metals in future applications under demanding mechanical conditions.

**One-Sentence Summary:** The results of our experiments reveal that the strength of metallic alloys can surpass the long-standing theoretical limit of $E/10$.

**Main Text:**

The pursuit of materials with ideal strength is a long-term goal of scientists working in the fields of materials science and mechanics who are interested in intrinsic properties. After the ground-breaking work by Frenkel (*6*), the ideal strength of materials was estimated by the well-known $\frac{E}{N}$ rule, in which $E$ is Young's modulus; $N$ is a constant with a value of approximately 10 (*7*). With the advances in nanofabrication and nanomechanical testing, strengths near the theoretical limit have been reported from experiments on nanosized crystalline quasi-brittle materials, including silicon nanowires ($E/7$) (*8*), carbon nanotubes ($E/10$) (*9*), graphene ($E/9$) (*10*), and diamond nanoneedles ($\sim E/10$) (*11, 12*). However, whether the measured strength of metallic materials can reach a comparable level remains unclear.

Unlike quasi-brittle materials, single-crystal and multi-grain metals deform plastically via dislocation mechanisms that are strongly affected by vacancies, impurities, twins, and grain boundaries (*7*). Thus, reducing the sample size alone does not necessarily ensure that metals achieve their ideal strengths. Richter *et al.* reported a strength value of $\sim E/25$ for single-crystal Cu nanowhiskers (*1*). Chen *et al.* reported a strength of $\sim E/18$ for single-crystal Pd nanowhiskers (*2*). Kim *et al.* revealed the failure mechanism of thin-layer twin formations for <110> Al nanowires with a strength of $\sim E/23$ (*3*). Researchers have also attempted to push the strength limit using ultrahigh strain rate tests where specimens deformed under the uniaxial strain condition. Under this condition, the spall strength is compared with the pressure wave (P-wave) modulus $M = \rho c^2$ in which $\rho$ and $c$ are the material density and the sound of speed, respectively. Coakley *et al.* conducted laser ablation on polycrystalline Cu foils at strain rates of $\dot{\varepsilon} \sim 5 \times 10^8$ s$^{-1}$ and reported a spall strength of $\sim M/25$ (*4*). Righi *et al.* reported high spall strengths for single-crystal ($\sim M/26$) and nanocrystals of Fe ($\sim M/42$) at strain rates of $\sim 2 \times 10^7$ s$^{-1}$ (*5*). Single-crystal and nanocrystalline metals undergo extensive plastic deformation mediated by dislocation movements (predominant at low or moderate strain rates (*13, 14*)) and mechanical twinning (predominant at ultrahigh strain rates (*5*)) prior to failure, which accounts for the notable difference between experimental and ideal strength values.

Metallic glasses (MGs) are renowned for their exceptionally high elastic limits and strengths due to their amorphous nature, which prevents the classic plastic deformation mechanisms (*15, 16*). Even under quasi-static loading, the mechanical strength of bulk MGs can reach $E/50$ (*17*). Subsequently, Tian *et al.* achieved strengths of $\sim E/20$ for Cu-Zr nanowires (*18*). This is because in nanosized MGs, it is difficult to launch shear banding, the main failure mechanism of MGs at low strain rates, due to the reduced number of clusters of shear transformation zones (STZs) (*18, 19*). Tang et al. conducted plate impact tests on MGs to achieve higher strain rates ($\sim 10^6$ s$^{-1}$) (*20*). The spalling at fracture surfaces of MGs showed cup-cone structures, indicating a mixed failure mechanism of cavitation and local shear banding. Nevertheless, Tang *et al.* obtained the spall strength of approximately $M/38$ (or $\sim E/21$). Thus, $E/20-E/25$ has long been regarded as the upper limit of the measurable strength for metals.

In contrast, the strength and the mechanisms of failure for MGs at extreme strain rates remain largely unexplored. In this study, we tested Cu$_{50}$Zr$_{50}$ at strain rates $\dot{\varepsilon} > 1\times 10^7$ s$^{-1}$ to assess its mechanical properties under these extreme conditions. The experimental observations were complemented with large-scale molecular dynamics (MD) simulations and continuum models to elucidate the unusual underlying mechanisms.

## Results

Thin $Cu_{50}Zr_{50}$ MG discs (100 μm thick) were fabricated using the single-roller melt-spinning method. We hit the front surface of each sample using a nanosecond Nd:glass laser (Shanghai Shenguang-II laser facility, National Laboratory on High Power Laser and Physics in Shanghai, China)to generate a shock (Fig. 1A). Two laser pulse durations (1 and 2.5 ns) and laser energy inputs ranging from 2 to 15 J were employed. A line image velocity interferometer system for any reflector (VISAR) was used to measure the rear free surface velocity (FSV) and deduce the spall strength and strain rate; a detailed analysis is given in Materials and Methods. Fig. S1A and B illustrate the representative FSV curves for the two pulse durations. The peak values of FSV ($v_{fsp}$), 870−1750 m/s, were significantly greater than those previously reported using the plate impact approach (300−600 m/s) (20-23). The tensile strain rates reached $1.4–2.8 \times 10^7\ s^{-1}$ (see Fig. S3). At high energy inputs, spallation occurred as the result of the interactions between incident and reflected shock waves, and the corresponding spall strengths varied from 5.6 to 9.8 GPa (Fig. 1B). The highest spall strength was approximately $M/13$ (the measured P-wave modulus $M = 126.9$ GPa, see Table S1). Note that $E = (1+v)(1-2v)M/(1-v)$, thus, taking Poisson's ratio $v \sim 0.4$ for $Cu_{50}Zr_{50}$ (24), the material strength approached approximately $E/6$. To the best of our knowledge, this was the first time that the measured strength of metals surpasses $E/10$ and approaches $M/10$. Notably, this record-breaking strength was achieved in the absence of the strengthening mechanisms present in crystalline metals, such as dislocation interactions and grain boundary strengthening. It is crucial to identify the underlying mechanism that grants the amorphous alloys exceptional mechanical strength.

First, we performed fractography on the spall plane. Scanning electron microscopy (SEM) images showed dimples rather than the cup-and-cone features on the spall plane (Fig. 2A–B and Fig. S6). More interestingly, the dimple size showed strong strain rate dependence (Fig. 2 and Fig. S6). At $\dot{\varepsilon} = 1.9 \times 10^7\ s^{-1}$ (5.0 J laser power and 2.5 ns pulse duration), the dimple diameter ranges from 2 to 3 μm. In contrast, at $\dot{\varepsilon} = 2.8 \times 10^7\ s^{-1}$ (14.3 J laser power and 2.5 ns pulse duration), a hierarchical dimple structure was seen; the majority of dimples had diameters of several hundred nanometers and only a small percentage exceeded one micron. Using a focused ion beam (FIB), we cut into the spall surfaces to examine the void distribution in the thickness direction. The specimen shocked by the 5.0 J laser pulse contained micro-voids scattered within a few microns beneath the spall surface (Fig. 3). In contrast, the specimen shocked by the 8.4 J laser pulse contained interconnected nanovoids underneath the spall surface.

These characterizations indicated that the material failed predominantly due to void growth and coalescence rather than shear banding. Intrinsic spatial heterogeneity is essential to their unique structure–property relationship (25-29). MGs consist of stable regions where atoms are more densely packed and rheological regions where atoms are more loosely packed, i.e., defective spots. At low and moderate strain rates, these defective spots distort and serve as nucleation sites for shear banding, i.e., shear transformation zones (STZs) (30). However, the laser pulse durations in our study, 1 and 2.5 ns, are substantially shorter than the timescale for shear band initiation, which is typically from several to tens of microseconds (31, 32). Thus, at extremely fast strain rates, these defective spots serve as cavitation nucleation sites, also known as the tension transformation zones or TTZs (33). Moreover, the higher the strain rate is, the greater the number of TTZs that are activated. This is because increasing the tensile hydrostatic stress decreases the free energy barrier for cavitation in MG, as suggested by Guan et al. (34).

Large-scale MD simulations of the uniaxial strain tension of $Cu_{50}Zr_{50}$ MG at strain rates in the range of $1 \times 10^8 - 5 \times 10^9$ s$^{-1}$ offer more insight into microscopic material failure mechanisms. First, the ultimate strength of the Cu-Zr MG increased with the strain rate, which was consistent with the results of our laser shock tests (Fig. 1B and Fig. S5C). However, the strain rate sensitivity of the material strength obtained from MD simulations was notably less than that found in our experiments, since the material strength approached its limit at $M/10$. In addition, as the strain rate increased, the number of voids increased, but the average void size decreased (Fig. 2C−D and Fig. S7). At relatively low strain rates, e.g., $1 \times 10^8$ s$^{-1}$, only a few isolated voids nucleated when the material reached its maximum stress. These voids grew independently as the material weakened and eventually failed. At high strain rates, e.g., $5 \times 10^9$ s$^{-1}$, however, the number of void nucleation sites increased sharply. The extremely close proximity of voids caused void growth and coalescence, which resulted in material failure.

Thus, our experiments and simulations helped us complete the failure mechanism diagram for metals shown in Fig. S8. At extreme strain rates, the nucleation, growth, and coalescence of voids become the dominant failure mechanisms for MGs. However, the void growth kinetics are fundamentally distinct from those of crystalline metals, in which dislocation movement and twinning play crucial roles. To establish the connection between void growth and rate-dependent spall strength, we adopted the Curran-Seaman-Shockey model to describe the growth of voids under the control of surface energy (35):

$$\frac{1}{D}\frac{dD}{dt} = \begin{cases} m(\sigma_h - \sigma_c), & \text{when } \sigma_h > \sigma_c \\ 0, & \text{when } \sigma_h \leq \sigma_c \end{cases}, \quad (1)$$

where $m$ is a mobility coefficient, $\sigma_h$ is the hydrostatic stress, and $\sigma_c$ is the critical/threshold hydrostatic stress. Assuming that void instability followed the classical nucleation theory, then the critical stress $\sigma_c = 4\gamma/D$ (36), in which $\gamma = 1.28$ J/m$^2$ is the surface energy obtained by MD simulations (see Materials and Methods). The hydrostatic stress in the uniaxial strain condition is $\sigma_h = (1+\nu)\sigma/[3(1-\nu)]$, where $\sigma = M\dot{\varepsilon}t$ is the normal stress in the thickness direction and $\nu = 0.41$ is the Poisson's ratio given by MD. Solving Eq. (1) yields the evolution of the void size (the details are given in the Supplementary Materials):

$$D(t) = \begin{cases} \left[D_0 \exp(-\chi^2 t_0^2) - \frac{2\sqrt{\pi}m\gamma}{\chi}(\text{erf}(\chi t) - \text{erf}(\chi t_0))\right]\exp(\chi^2 t^2), & t > t_0 \\ D_0, & t \leq t_0 \end{cases}, \quad (2)$$

where $D_0$ is the initial void diameter, $\chi = \sqrt{\frac{M\dot{\varepsilon}m(1+\nu)}{6(1-\nu)}}$ and $t_0 = \frac{12\gamma(1-\nu)}{D_0 M\dot{\varepsilon}(1+\nu)}$. Setting the mobility coefficient $m = 25$ (Pa·s)$^{-1}$ and $D_0 \sim 0.4$ nm, we noted remarkable agreement between the void evolution curves predicted by Eq. (2) and the void growth trends from MD simulations for all strain rates (Fig. S9); these results validated our model.

Despite the fact that the material strength increased with the strain rate in our MD simulations, we noted that the critical void diameter (when all curves reached their peak stresses) remained nearly constant ($D_c \approx 7.3$ nm), as shown in Fig. S5D. This constant critical void diameter stemmed from the onset of mechanical instability due to competition between surface energy and strain energy; the details of the derivations are shown in the Supplementary Materials. Therefore, we used $D(t_c) = 7.3$ nm as the criterion to estimate the spall strength ($\sigma_s = M\dot{\varepsilon}t_c$) for the specimens in the laser shock tests. When we employed a smaller mobility coefficient $m = 0.3$ (Pa·s)$^{-1}$ and $D_0 \sim 2$ nm (*37*) and kept all the parameters the same as those used in the MD results, our model predicted strain rate dependence of the spall strength that was in excellent agreement with our experiments ($1 \times 10^7$ s$^{-1} < \dot{\varepsilon} < 3 \times 10^7$ s$^{-1}$) and previous reports ($1 \times 10^5$ s$^{-1} < \dot{\varepsilon} < 1 \times 10^6$ s$^{-1}$), as shown in Fig. 1B. Notably, mobility coefficient used for real materials was much lower than that used for MD simulations. This was due to the substantial difference between the energy states of the in silico model and the real material. Wang *et al*. emphasized that the substantially faster cooling rates in in silico MG models than those for practical MGs prepared by melt-spinning resulted in a far smaller activation energy for atomic motion and, thus, significantly greater mobility (*38*).

Although the experiments by Coakley *et al*. and Righi *et al*. reached strain rates similar to those in this study (*4, 5*), the tested polycrystalline or single crystal metals exhibited spall strengths < *M*/20. This is because the plastic deformation of crystalline metals is mediated mainly by dislocation and twinning activities. Therefore, the void growth model proposed by Wilkerson and Ramesh (*39, 40*) is suitable for their cases:

$$\frac{1}{D}\frac{dD}{dt} = \begin{cases} \frac{1}{3}bn_m c_s \tanh\left[\frac{3}{4}\frac{b}{Bc_s}(\sigma_h - \sigma_c)\right], \text{ when } \sigma_h > \sigma_c \\ 0, \text{ when } \sigma_h \leq \sigma_c \end{cases} \quad (3)$$

where $n_m$ is the mobile dislocation density, $b$ is the Burgers vector, $c_s$ is the shear wave speed and $B$ is the drag coefficient. Taking the parameters for copper, for instance, $n_m = 2 \times 10^{17}$ m$^{-2}$ (*13*), $b = 0.25$ nm, $c_s = 2469$ m/s, and $B = 1.6 \times 10^{-5}$ Pa·s (*41*), we estimated the void growth rate $\dot{D}$ from 40 to 400 m/s under an overpressure $(\sigma_h - \sigma_c)$ of approximately between 1 and 4 GPa. This estimated rate was in good agreement with a recent experimental characterization Cu (from 50 to 680 m/s) (*4*). In the absence of dislocation-mediated mechanisms, the growth rate in Cu-Zr MG is only approximately from 0.3 to 12 m/s, nearly two orders of magnitude slower than in crystalline metals. This much lower void growth rate endows our material with exceptional spall strength. Fig. 4 summarizes the state-of-the-art experimental measurements of the ultimate strength of various metallic materials; our measurements advance the record for strength to ~*M*/13.

In summary, we conducted shock experiments on Cu-Zr MG at ultrahigh strain rates near the capacity of MD simulations by employing nanosecond laser pulses. Our study raised the measured strength of metallic materials to the unprecedented level of *M*/13 and approached the theoretical limit. Due to inadequate time for the development of shear banding at strain rates faster than $1.0 \times 10^7$ s$^{-1}$ (loading timescale less than 5 ns) in our experiments, the material failed predominantly due to collective void nucleation and growth. Large-scale MD simulations and strain-rate-dependent hierarchical void structures on the spall surfaces indicated that faster strain rates activated a greater number of TTZs in MGs. When the voids reached a critical size, the material weakened due to mechanical instability from competition between surface energy and strain energy. We demonstrated that a void growth model governed by surface energy accurately depicted the strain rate dependence of the spall strength. In this study, the mechanical properties and failure mechanisms revealed by the extremely fast mechanical loading conditions enhanced our understanding of the time-dependent behavior of amorphous solids. Our findings also provided new prospects for utilizing amorphous phases to optimize the performance and design of metallic materials for applications under extremely fast mechanical conditions. Follow-up research should concentrate on competition between shear banding and cavitation instability in MGs under extreme conditions, as cavitation has been largely overlooked in current plasticity theories for disordered materials (*42*).

**References and Notes**

1. G. Richter *et al.*, Ultrahigh strength single crystalline nanowhiskers grown by physical vapor deposition. *Nano Letters* **9**, 3048 (2009).
2. L. Y. Chen, M. R. He, J. Shin, G. Richter, D. S. Gianola, Measuring surface dislocation nucleation in defect-scarce nanostructures. *Nat Mater* **14**, 707-713 (2015).
3. S.-H. Kim *et al.*, Deformation twinning of ultrahigh strength aluminum nanowire. *Acta Materialia* **160**, 14-21 (2018).
4. J. Coakley, A. Higginbotham, D. McGonegle, J. Wark, Femtosecond quantification of void evolution during rapid material failure. *Science Advances* **6**, eabb4434 (2020).
5. G. Righi *et al.*, Towards the ultimate strength of iron: spalling through laser shock. *Acta Materialia*, (2021).
6. J. Frenkel, Zur theorie der elastizitätsgrenze und der festigkeit kristallinischer körper. *Zeitschrift für Physik* **37**, 572-609 (1926).
7. T. Zhu, J. Li, Ultra-strength materials. *Progress in Materials Science* **55**, 710-757 (2010).
8. H. Zhang *et al.*, Approaching the ideal elastic strain limit in silicon nanowires. *Science Advances* **2**, e1501382 (2016).
9. B. Peng *et al.*, Measurements of near-ultimate strength for multiwalled carbon nanotubes and irradiation-induced crosslinking improvements. *Nat Nanotechnol* **3**, 626-631 (2008).
10. C. Lee, X. Wei, J. W. Kysar, J. Hone, Measurement of the Elastic Properties and Intrinsic Strength of Monolayer Graphene. *Science* **321**, 385-388 (2008).
11. A. Nie *et al.*, Approaching diamond's theoretical elasticity and strength limits. *Nature communications* **10**, 1-7 (2019).
12. C. Dang *et al.*, Achieving large uniform tensile elasticity in microfabricated diamond. *Science* **371**, 76-78 (2021).
13. E. M. Bringa, S. Traiviratana, M. A. Meyers, Void initiation in fcc metals: Effect of loading orientation and nanocrystalline effects. *Acta Materialia* **58**, 4458-4477 (2010).
14. M. A. Meyers, A. Mishra, D. J. Benson, Mechanical properties of nanocrystalline materials. *Progress in Materials Science* **51**, 427-556 (2006).
15. A. L. Greer, E. Ma, Bulk metallic glasses: at the cutting edge of metals research. *MRS bulletin* **32**, 611-619 (2007).
16. M. Chen, Mechanical behavior of metallic glasses: microscopic understanding of strength and ductility. *Annu. Rev. Mater. Res.* **38**, 445-469 (2008).
17. W. H. Wang, Elastic moduli and behaviors of metallic glasses. *Journal of Non-Crystalline Solids* **351**, 1481-1485 (2005).
18. L. Tian *et al.*, Approaching the ideal elastic limit of metallic glasses. *Nat Commun* **3**, 609 (2012).
19. J. R. Greer, J. T. M. De Hosson, Plasticity in small-sized metallic systems: Intrinsic versus extrinsic size effect. *Progress in Materials Science* **56**, 654-724 (2011).

20. X. C. Tang *et al.*, Cup-cone structure in spallation of bulk metallic glasses. *Acta Materialia* **178**, 219-227 (2019).
21. G. Ding *et al.*, Ultrafast extreme rejuvenation of metallic glasses by shock compression. *Science Advances* **5**, (2019).
22. J. P. Escobedo, Y. M. Gupta, Dynamic tensile response of Zr-based bulk amorphous alloys: Fracture morphologies and mechanisms. *Journal of Applied Physics* **107**, (2010).
23. S. Zhuang, J. Lu, G. Ravichandran, Shock wave response of a zirconium-based bulk metallic glass and its composite. *Applied Physics Letters* **80**, 4522-4524 (2002).
24. W. L. Johnson, K. Samwer, A universal criterion for plastic yielding of metallic glasses with a (T/Tg) 2/3 temperature dependence. *Phys Rev Lett* **95**, 195501 (2005).
25. Z. Lu, W. Jiao, W. H. Wang, H. Y. Bai, Flow unit perspective on room temperature homogeneous plastic deformation in metallic glasses. *Phys Rev Lett* **113**, 045501 (2014).
26. J. Ding *et al.*, Universal structural parameter to quantitatively predict metallic glass properties. *Nat Commun* **7**, 13733 (2016).
27. C. Liu, R. Maaß, Elastic Fluctuations and Structural Heterogeneities in Metallic Glasses. *Advanced Functional Materials* **28**, (2018).
28. F. Zhu, S. Song, K. M. Reddy, A. Hirata, M. Chen, Spatial heterogeneity as the structure feature for structure-property relationship of metallic glasses. *Nat Commun* **9**, 3965 (2018).
29. J. C. Qiao *et al.*, Structural heterogeneities and mechanical behavior of amorphous alloys. *Progress in Materials Science* **104**, 250-329 (2019).
30. A. L. Greer, Y. Q. Cheng, E. Ma, Shear bands in metallic glasses. *Materials Science and Engineering: R: Reports* **74**, 71-132 (2013).
31. M. Seleznev, I. Yasnikov, A. Vinogradov, On the shear band velocity in metallic glasses: A high-speed imaging study. *Materials Letters* **225**, 105-108 (2018).
32. P. Thurnheer, F. Haag, J. F. Löffler, Time-resolved measurement of shear-band temperature during serrated flow in a Zr-based metallic glass. *Acta Materialia* **115**, 468-474 (2016).
33. M. Q. Jiang, Z. Ling, J. X. Meng, L. H. Dai, Energy dissipation in fracture of bulk metallic glasses via inherent competition between local softening and quasi-cleavage. *Philosophical Magazine* **88**, 407-426 (2008).
34. P. Guan, S. Lu, M. J. Spector, P. K. Valavala, M. L. Falk, Cavitation in amorphous solids. *Phys Rev Lett* **110**, 185502 (2013).
35. L. Seaman, D. R. Curran, D. A. Shockey, Computational models for ductile and brittle fracture. *Journal of Applied Physics* **47**, 4814-4826 (1976).
36. F. Abraham, *Homogeneous nucleation theory: the pretransition theory of vapor condensation*. (Elsevier, 2012), vol. 1.
37. J. Li, F. Spaepen, T. C. Hufnagel, Nanometre-scale defects in shear bands in a metallic glass. *Philosophical Magazine A* **82**, 2623-2630 (2002).
38. Y.-J. Wang, J.-P. Du, S. Shinzato, L.-H. Dai, S. Ogata, A free energy landscape perspective on the nature of collective diffusion in amorphous solids. *Acta Materialia* **157**, 165-173 (2018).
39. J. W. Wilkerson, K. T. Ramesh, A dynamic void growth model governed by dislocation kinetics. *Journal of the Mechanics and Physics of Solids* **70**, 262-280 (2014).
40. J. W. Wilkerson, On the micromechanics of void dynamics at extreme rates. *International Journal of Plasticity* **95**, 21-42 (2017).
41. H. Fan, Q. Wang, J. A. El-Awady, D. Raabe, M. Zaiser, Strain rate dependency of dislocation plasticity. *Nat Commun* **12**, 1845 (2021).

42. Z. Budrikis, D. F. Castellanos, S. Sandfeld, M. Zaiser, S. Zapperi, Universal features of amorphous plasticity. *Nat Commun* **8**, 15928 (2017).
43. G. Wu, K. C. Chan, L. Zhu, L. Sun, J. Lu, Dual-phase nanostructuring as a route to high-strength magnesium alloys. *Nature* **545**, 80-83 (2017).
44. C. C. Wang *et al.*, Sample size matters for Al88Fe7Gd5 metallic glass: Smaller is stronger. *Acta Materialia* **60**, 5370-5379 (2012).
45. Y. Zhu *et al.*, Size effects on elasticity, yielding, and fracture of silver nanowires:In situ experiments. *Physical Review B* **85**, (2012).
46. H. Shu *et al.*, Plastic behavior of aluminum in high strain rate regime. *Journal of Applied Physics* **116**, (2014).

**Acknowledgments:**

The experiment was performed using Shenguang-II Nd:glass high-power laser facility at the National Laboratory on High Power Laser and Physics in Shanghai, China.

**Funding:** The authors greatly appreciate the support by the National Natural Science Foundation of China (Grant Nos. 12172005, 11890681, and 11988102), and the National Key R&D Program of China (Grant No. 2020YFE0204200).

**Author contributions:** W. Z. and H. S. both carried out the experiments and analyzed the results. W. Z. and Z. L. developed and validated the theoretical model. Z. L. performed the MD simulations and data analyses. X. Wei and H. Gao conceived the concept and supervised the study. All authors contributed to manuscript writing.

**Competing interests:** All other authors declare they have no competing interests.

**Data and materials availability:** All data are available in the main text or the supplementary materials. Additional data related to this paper may be requested from the authors.

**Supplementary Materials**

Materials and Methods

Supplementary Text

Figs. S1 to S9

Tables S1

References (*46–50*)

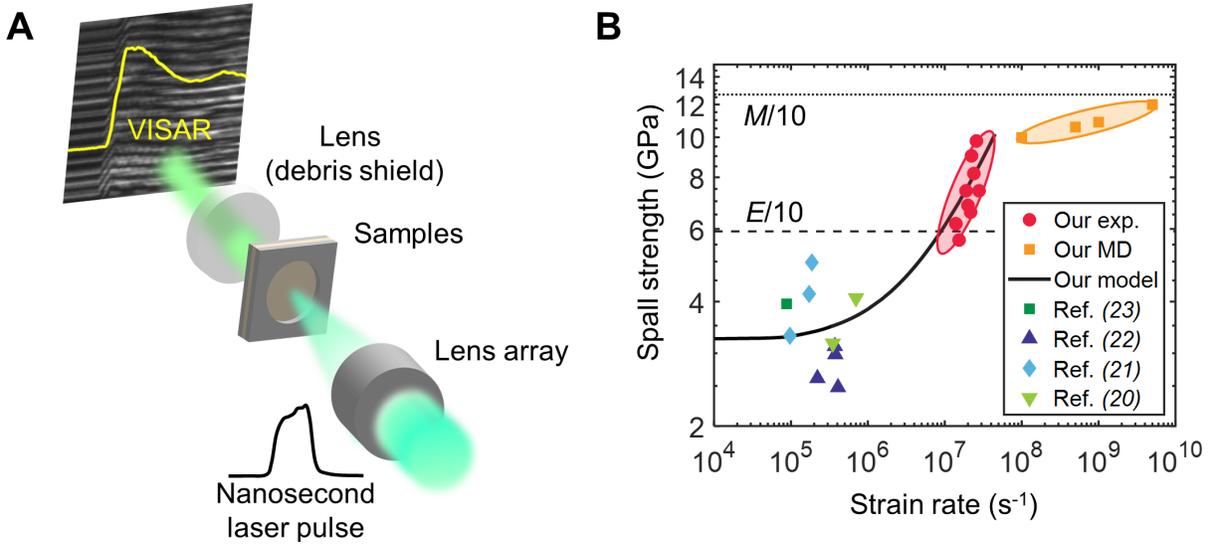

**Fig. 1. Laser-induced shock tests of Cu-Zr MG disks.** (**A**) Schematic diagram of the laser-induced shock experimental equipment. (**B**) Summary of data for spall strength vs. tensile strain rate. The results for Zr-based MGs from previous reports are included for comparison (Refs. (*20-23*)). The solid line is the prediction of the rate-dependent strength based on our kinetic model of void growth (Eqs. (1) and (2)). The dotted line and the dashed line indicate the limits of $M/10$ and $E/10$, respectively.

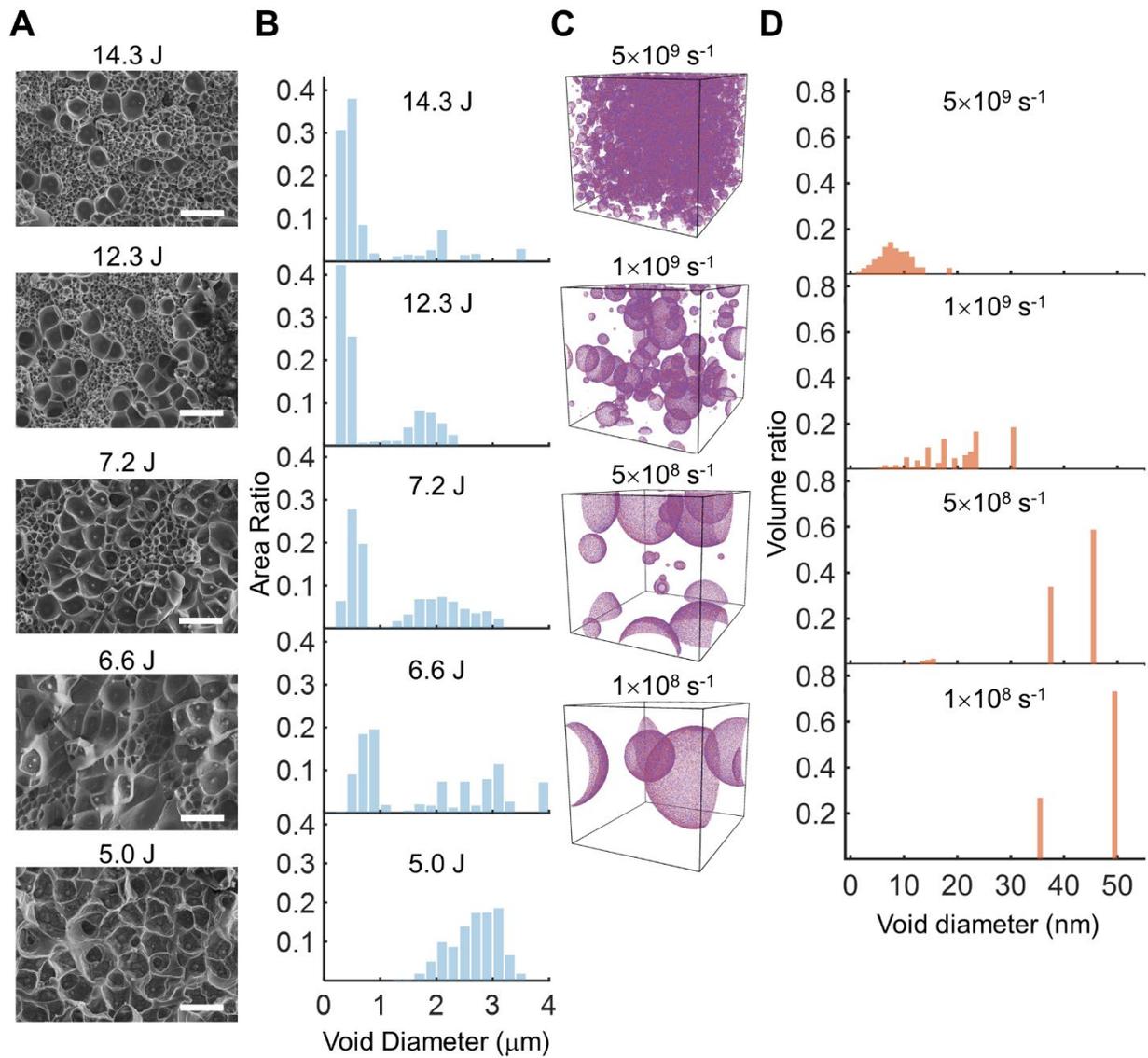

**Fig. 2. Rate-dependent void size distribution on the spall plane.** (**A**) SEM micrographs of the spall surfaces of the samples tested by the laser with 2.5 ns pulse duration and different input energies (from bottom to top: 5.0 J, 6.6 J, 7.2 J, 12.3 J, and 14.3 J). Scale bar: 5 μm. (**B**) Histograms of the void size distributions at different laser powers, i.e., strain rates. (**C**) Final void morphology obtained from large-scale MD simulations for $Cu_{50}Zr_{50}$ stretched at various strain rates. (**D**) The corresponding statistics for the void sizes obtained in MD simulations.

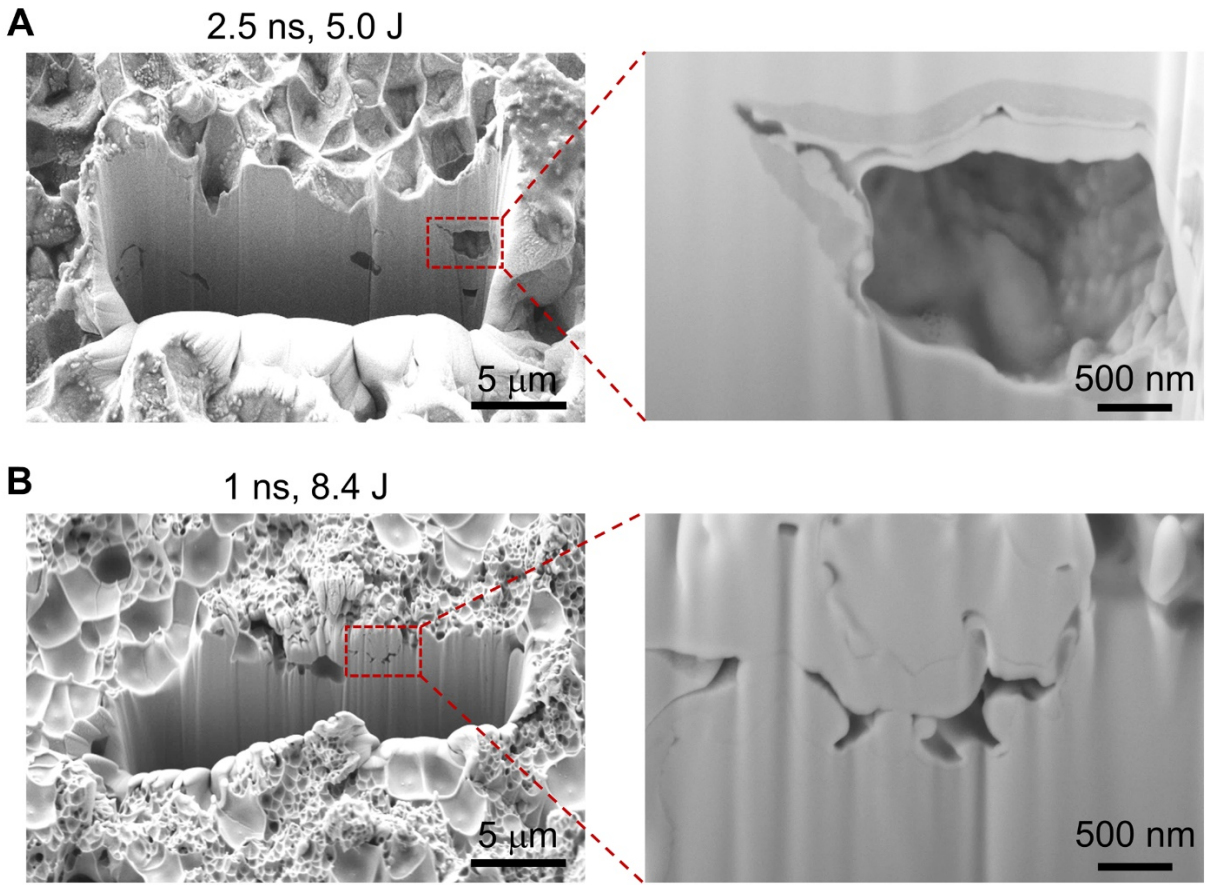

**Fig. 3. Rate-dependent void distributions underneath the spall fracture surface**. SEM micrographs on the sidewall of the well milled by FIB show void growth and coalescence underneath the spall planes for the sample tested by the laser with 2.5 ns pulse duration and 5.0 J energy (**A**) and the laser with 1 ns pulse duration and 8.4 J energy (**B**).

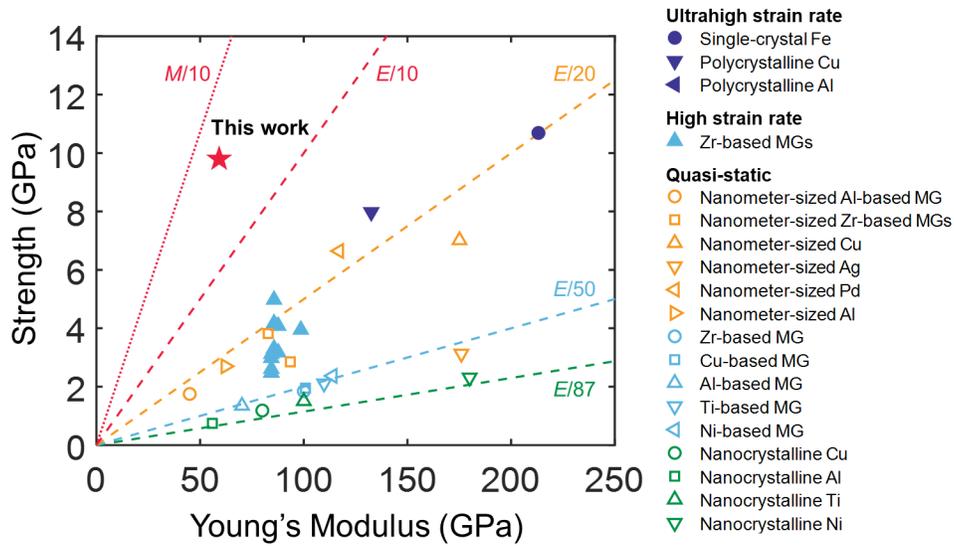

**Fig. 4. Summary of the measured ultimate strengths for metallic materials.** Representative experimental results for crystalline and amorphous metals tested under quasi-static conditions (*1-3, 18, 43-45*), at high rates ($10^4$ s$^{-1}$ < $\dot{\varepsilon}$ < $10^6$ s$^{-1}$) (*20-23*), and ultrahigh rates ($\dot{\varepsilon}$ > $10^6$ s$^{-1}$) (*4, 5, 46*) are included for comparison. Our study on Cu-Zr MG shows the record-setting spall strength of ~*M*/13 for metals.

# Supplementary Materials for

## Amorphous alloys surpass E/10 strength limit at extreme strain rates

Wenqing Zhu[*], Zhi Li[*], Hua Shu, Huajian Gao[†], Xiaoding Wei[†]

[*]These authors contributed equally to this work.

[†]To whom correspondence and requests for materials should be addressed. E-mail: huajian.gao@ntu.edu.sg (H.G.); xdwei@pku.edu.cn (X.W.)

**This PDF file includes:**

Materials and Methods
Supplementary Text
Figs. S1 to S9
Tables S1

## Materials and Methods

Laser-induced Shock Experiments

Shock experiments were performed using the Shenguang-II Nd:glass laser facility (converted to a wavelength of 351 nm) at the National Laboratory for High Power Lasers and Physics in Shanghai, China. The temporal profile of the laser pulse was approximately square, and pulse durations (full width at half maximum) of 1 ns and 2.5 ns were adopted. A lens array (LA) was used to eliminate the large-scale spatial modulation and obtain a flat-topped profile in the focal plane. The optical system (lens + LA) had a focal spot over a flat region of dimensions ~0.5 × 0.5 mm$^2$. Laser energy in the range of 2–15 J was chosen so that spallation could occur without the laser completely plasmarizing the samples. Time-resolved FSV profiles of the shocked samples were measured with VISAR. The time window and resolution of the VISAR system were 20 ns and 50 ps, respectively.

Laser shock experiments on the MG samples with a step were used to measure the longitudinal speed of sound (Fig. S2A) (*46*). The time delay $t_d$ between FSV signals for two surfaces was measured. In this way, the longitudinal speed of sound $c = h/t_d = 4487$ m/s was obtained; it was in good agreement with the value obtained from molecular dynamics simulations ($c = 4340$ m/s) (*47*). The material density $\rho = 6.30$ g/cm$^3$ was measured using the Archimedes method. The amplitude of the compressive stress of a shock wave was calculated as $\sigma_p = \rho c v_{fsp}/2$, where $v_{fsp}$ is the peak value of FSV. The spall strength was calculated using $\sigma_s = \rho c \Delta v_{fs}/2$, where $\Delta v_{fs}$ and $\Delta t$ are the velocity and time difference, respectively, as shown in Fig. S2B. Accordingly, the tensile strain rate was obtained by $\dot{\varepsilon} = \Delta v_{fs} / (2c\Delta t)$.

Sample Preparation and Characterization of the Spall Planes

Cu$_{50}$Zr$_{50}$ MG samples with in-plane dimensions of 2 × 2 mm$^2$ and thickness of approximately 100 μm were prepared for laser shock experiments using the single-roller melt spinning method. X-ray diffraction (Empyrean XRD, Malvern Panalytical Ltd) was performed to verify the amorphous state of the samples (Fig. S4). After laser shock tests, the MG samples were retrieved, and microscopy (Hitachi FE-SEM S4800) was used to characterize the plane where spalling occurred. Then, the void distribution underneath the spall plane was characterized using a focused ion beam (FIB) to mill a rectangular well in the sample (ZEISS Crossbeam 340).

Molecular Dynamics Simulations

Molecular dynamics simulations were conducted to explore the cavitation kinetics of Cu$_{50}$Zr$_{50}$ MG under ultrahigh strain rates. The simulations were performed using LAMMPS open source code (*48*); a Finnis-Sinclair-type interatomic potential developed by Mendelev *et al*. was employed (*49*). In all the simulations, periodic boundary conditions were applied to all three axes. First, the simulation box containing 16 million randomly distributed Cu atoms and 16 million Zr atoms was heated to 2000 K and held for 200 ps. Then, the system was cooled to 300 K at a rate of $1.7 \times 10^{12}$ K/s. After that, the system equilibrated at 300 K for 200 ps. In all these processes, an isothermal-isobaric (NPT) ensemble was employed, and a time step of 1 fs was used. Finally, this simulation gave an atomistic model for Cu$_{50}$Zr$_{50}$ MG with dimensions of approximately 80 × 80 × 80 nm$^3$. The glassy state of the model was characterized by the pair distribution function shown in Fig. S5B. The glass transition temperature was approximately 750 K and determined by the change in the slopes of the potential energy vs. temperature curve during quenching (*50*).

Then, uniaxial strain tests were performed. The model was stretched along the x-axis to 20% strain at nominal strain rates ranging from $1 \times 10^8$ s$^{-1}$ to $5 \times 10^9$ s$^{-1}$. Meanwhile, the dimensions of the box along the y-axis and z-axis were fixed to ensure that the model deformed under uniaxial strain conditions. The microcanonical (NVE) ensemble was employed, and a time step of 1 fs was used in the mechanical tests.

To calculate the surface energy of $Cu_{50}Zr_{50}$ MG, two quenched samples were equilibrated at 300 K with a time step of 1 fs using NPT ensembles. One of the systems had periodic boundary conditions along all three axes, while the other system had periodic boundary conditions in the y- and z-directions and free surface boundary conditions in the x-direction. The total potential energy of the two systems averaged over 20 ps was used in the surface energy calculation.

**Supplementary Text**

Derivations of the analytical solution of the kinetic model of void growth

Under the control of surface energy, the evolution of the void diameter $D(t)$ based on Eq. (1) can be reorganized as follows:

$$\dot{D} = \begin{cases} \dfrac{1+v}{1-v}\dfrac{mM\dot{\varepsilon}t}{3}D - 4m\gamma, & t \geq t_0 \\ 0, & t > t_0 \end{cases}, \quad (S1)$$

where $t_0 = \dfrac{12\gamma(1-v)}{D_0 M\dot{\varepsilon}(1+v)}$ is the time when the void starts to grow, given by the condition $\sigma_h = 4\gamma/D_0$; $D_0$ is the initial void diameter. The following form can be used for the solution to the above equation:

$$D(t) = \begin{cases} f(t)\exp\left[\dfrac{m(1+v)}{6(1-v)}M\dot{\varepsilon}\left(t^2 - t_0^2\right)\right], & t \geq t_0 \\ D_0, & t > t_0 \end{cases}. \quad (S2)$$

Thus, for $t \geq t_0$, we have

$$\dot{D} = \dfrac{m(1+v)}{3(1-v)}M\dot{\varepsilon}tD + f'(t)\exp\left[\dfrac{m(1+v)}{6(1-v)}M\dot{\varepsilon}\left(t^2 - t_0^2\right)\right] = \dfrac{m(1+v)}{3(1-v)}M\dot{\varepsilon}tD - 4m\gamma, \quad (S3)$$

which leads to the analytical expression for $f(t)$:

$$\begin{aligned} f(t) &= D_0 - 4m\gamma \exp\left[\dfrac{m(1+v)}{6(1-v)}M\dot{\varepsilon}t_0^2\right]\int_{t_0}^{t}\exp\left[-\dfrac{m(1+v)}{6(1-v)}M\dot{\varepsilon}x^2\right]dx \\ &= D_0 - 2\gamma\sqrt{\dfrac{6\pi m(1-v)}{M\dot{\varepsilon}(1+v)}}\exp\left[\dfrac{m(1+v)}{6(1-v)}M\dot{\varepsilon}t_0^2\right] \times \\ &\quad \left[\text{erf}\left(\sqrt{\dfrac{M\dot{\varepsilon}m(1+v)}{6(1-v)}}t\right) - \text{erf}\left(\sqrt{\dfrac{M\dot{\varepsilon}m(1+v)}{6(1-v)}}t_0\right)\right] \end{aligned}. \quad (S4)$$

Notably, the initial condition $D(t_0) = D_0$ has been applied above. Finally, we obtain the analytical solution for the void diameter evolution $D(t)$:

$$D(t) = \begin{cases} \left\{ D_0 \exp\left(-\frac{m}{6}\frac{1+v}{1-v}M\dot{\varepsilon}t_0^2\right) - 2\gamma\sqrt{\frac{6\pi m(1-v)}{M\dot{\varepsilon}(1+v)}} \left[ \text{erf}\left(\sqrt{\frac{M\dot{\varepsilon}m(1+v)}{6(1-v)}}t\right) \right. \right. \\ \left. \left. -\text{erf}\left(\sqrt{\frac{M\dot{\varepsilon}m(1+v)}{6(1-v)}}t_0\right) \right] \right\} \exp\left(\frac{m}{6}\frac{1+v}{1-v}M\dot{\varepsilon}t^2\right), \ t > t_0 \\ D_0, \ t \le t_0 \end{cases} \quad \text{(S5)}$$

Critical void size from the mechanical instability analysis

The total energy of the system with a void of diameter $D$ is $F = \pi D^2\gamma - \pi D^3\psi/6$, in which the first term is the surface energy, and the second term is the elastic strain energy, where $\psi$ is the elastic strain energy density. Thus, the mechanical stability of the system breaks down when $dF/dD = 0$, which yields:

$$2D\gamma - \frac{D^2\psi}{2} = 0. \quad \text{(S6)}$$

Therefore, the critical void diameter at the instability is $D_c = 4\gamma/\psi$. The elastic strain energy density is approximately $\psi = M\varepsilon^2/2$ where $M = 126.9$ GPa is the P-wave modulus and $\varepsilon \sim 0.1$. Taking $\gamma = 1.28$ J/m$^2$ obtained from MD simulations, we estimate the critical void diameter $D_c \approx$ 8.1 nm, which agrees remarkably well with the value of $D_c \approx 7.3$ nm obtained from the MD simulations.

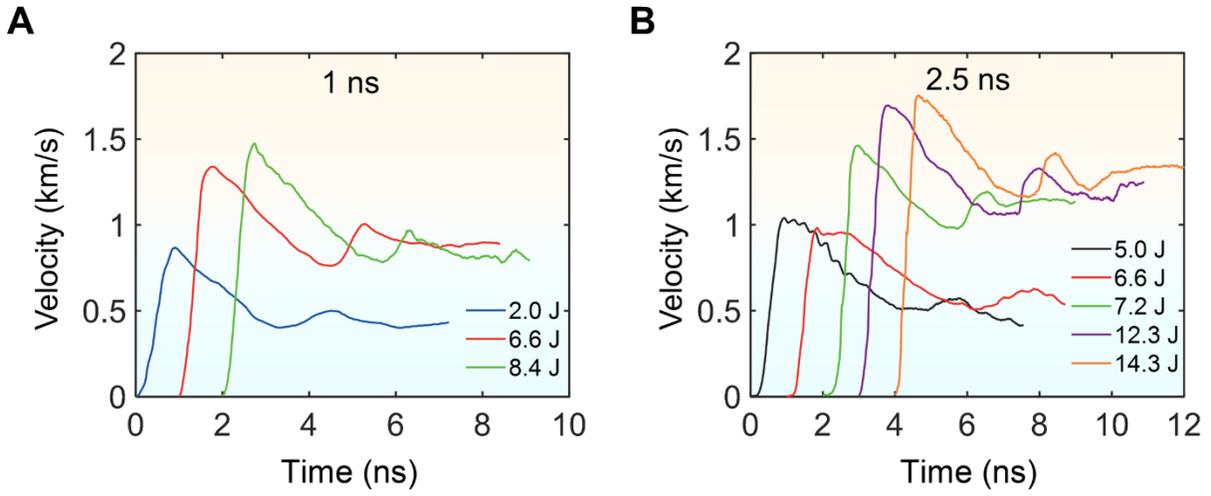

**Fig. S1.**
Time records of the free surface velocities for samples tested by laser pulses with durations of 1 ns (A) and 2.5 ns (B).

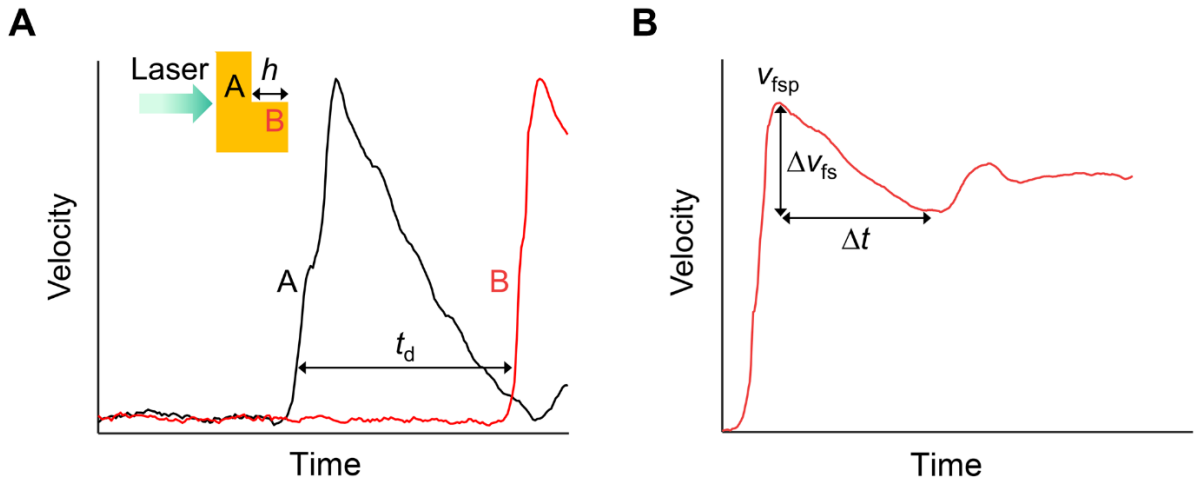

**Fig. S2.**
(A) Schematic diagram of the measurement of the longitudinal speed of sound by shock tests on samples with a step on the rear surface. (B) Typical FSV curve used to extract the spall strength and tensile strain rate.

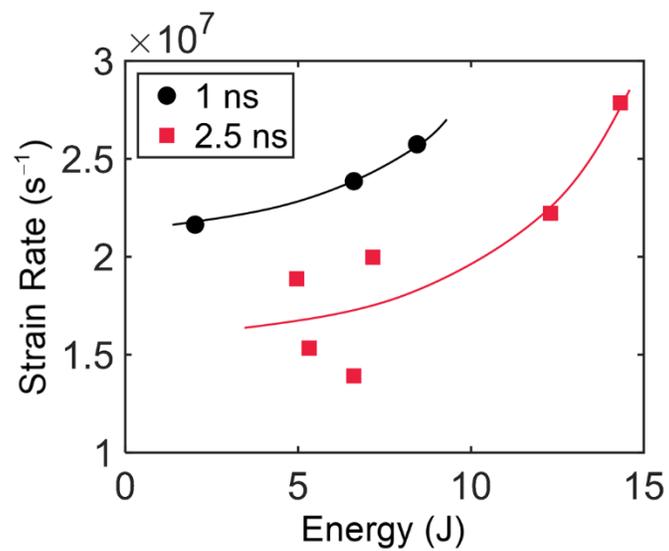

**Fig. S3.**
Tensile strain rate vs. input energy. The tensile strain rate increases with increasing laser power where either the energy rises or the duration becomes shorter.

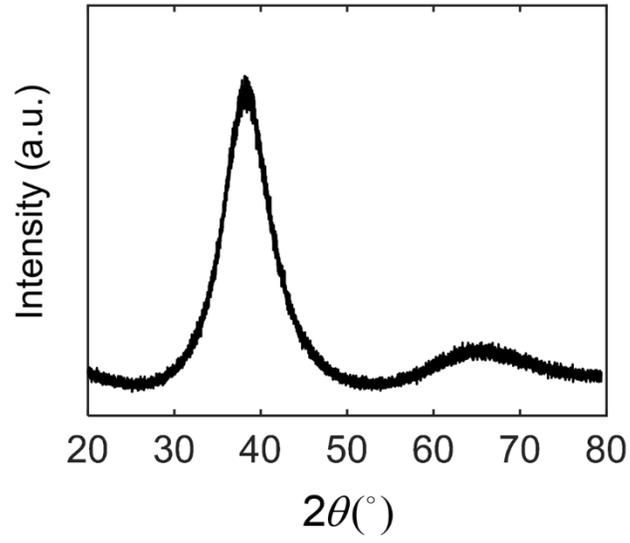

**Fig. S4.**
X-ray diffraction (XRD) spectrum of the as-cast MG samples that confirms the glassy state.

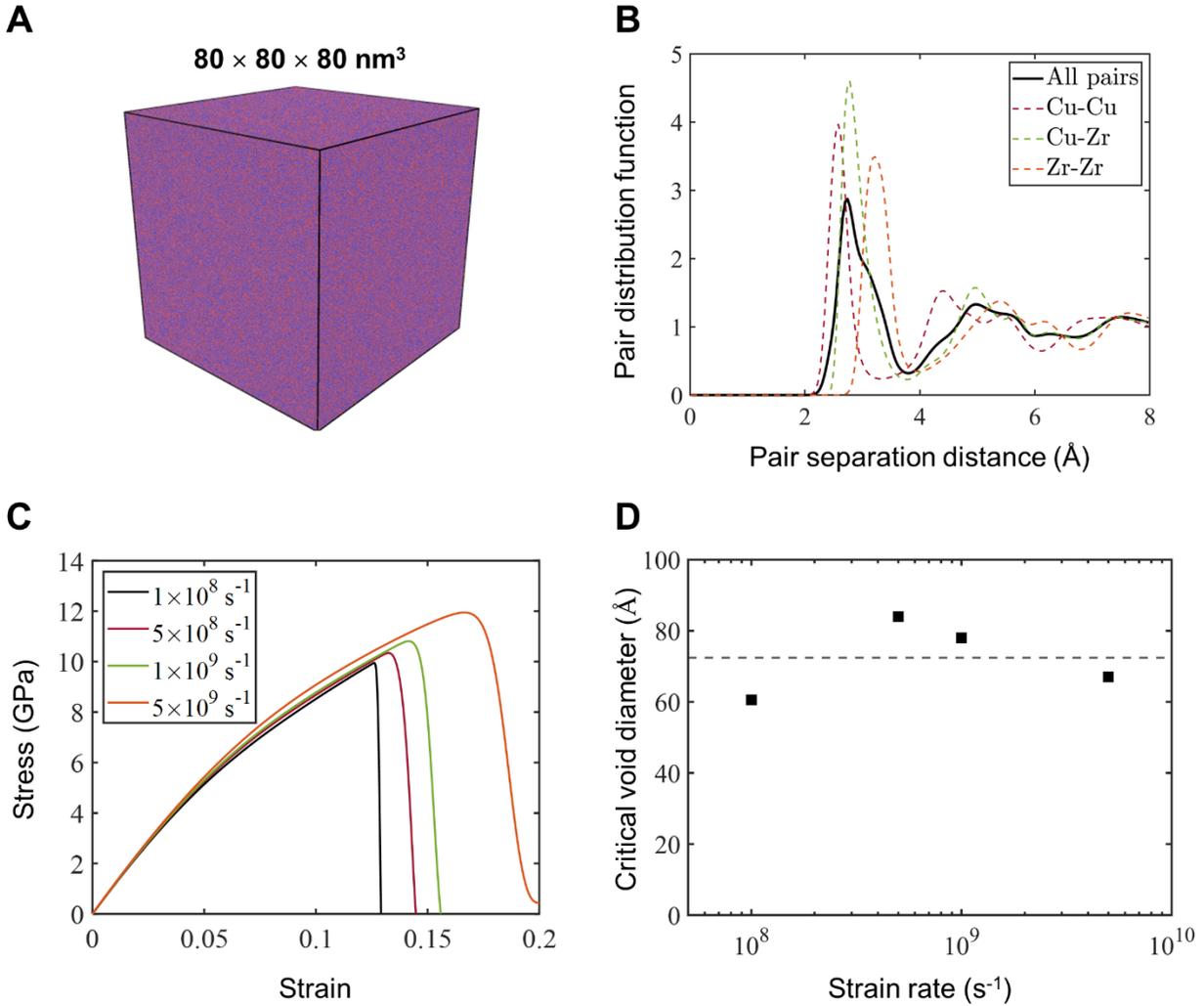

**Fig. S5.**
Large-scale molecular dynamics simulations of the uniaxial strain tensile tests on $Cu_{50}Zr_{50}$. (**A**) The $Cu_{50}$ $Zr_{50}$ model with dimensions of 80×80×80 nm³ used for the simulations. Red atoms represent Cu, and purple atoms represent Zr. (**B**) The radial pair distribution function for the CuZr MG model. (**C**) Stress–strain curves for the $Cu_{50}$ $Zr_{50}$ MG model at strain rates ranging from $1 \times 10^8$ s⁻¹ to $5 \times 10^9$ s⁻¹ under uniaxial tensile strain. (**D**) When the model is at its peak stress, the maximum void diameters are approximately 7.3 nm at different strain rates.

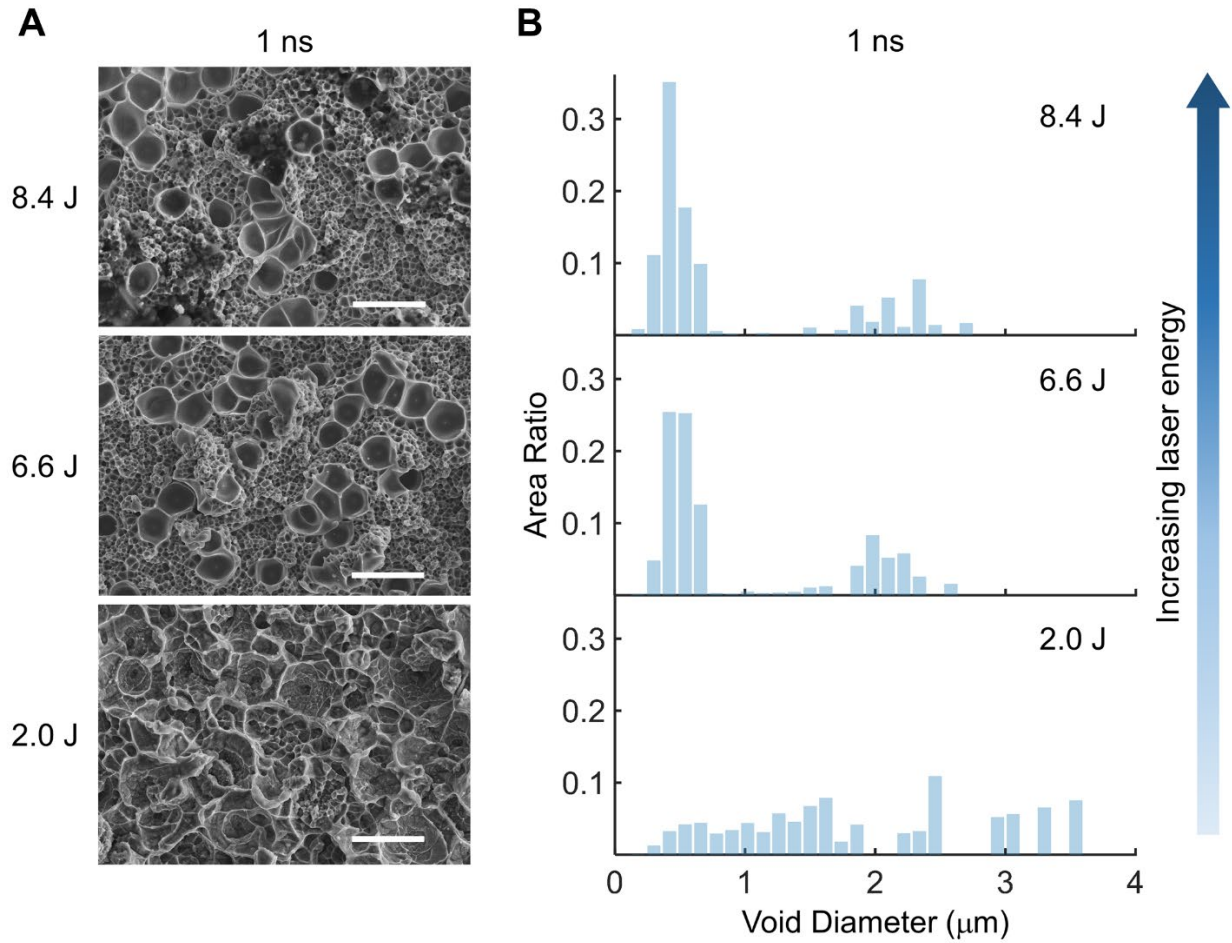

**Fig. S6.**
SEM micrographs and statistics of the void distributions on the spall surface for different laser input energies. (**A**) SEM micrographs of the spall surfaces of the samples tested by the laser with 1 ns duration but different input energies (from bottom to top: 2.0 J, 6.6 J, and 8.4 J). Scale bar, 5 μm. (**B**) Histograms show the statistics of the void diameters in the micrographs in (**A**).

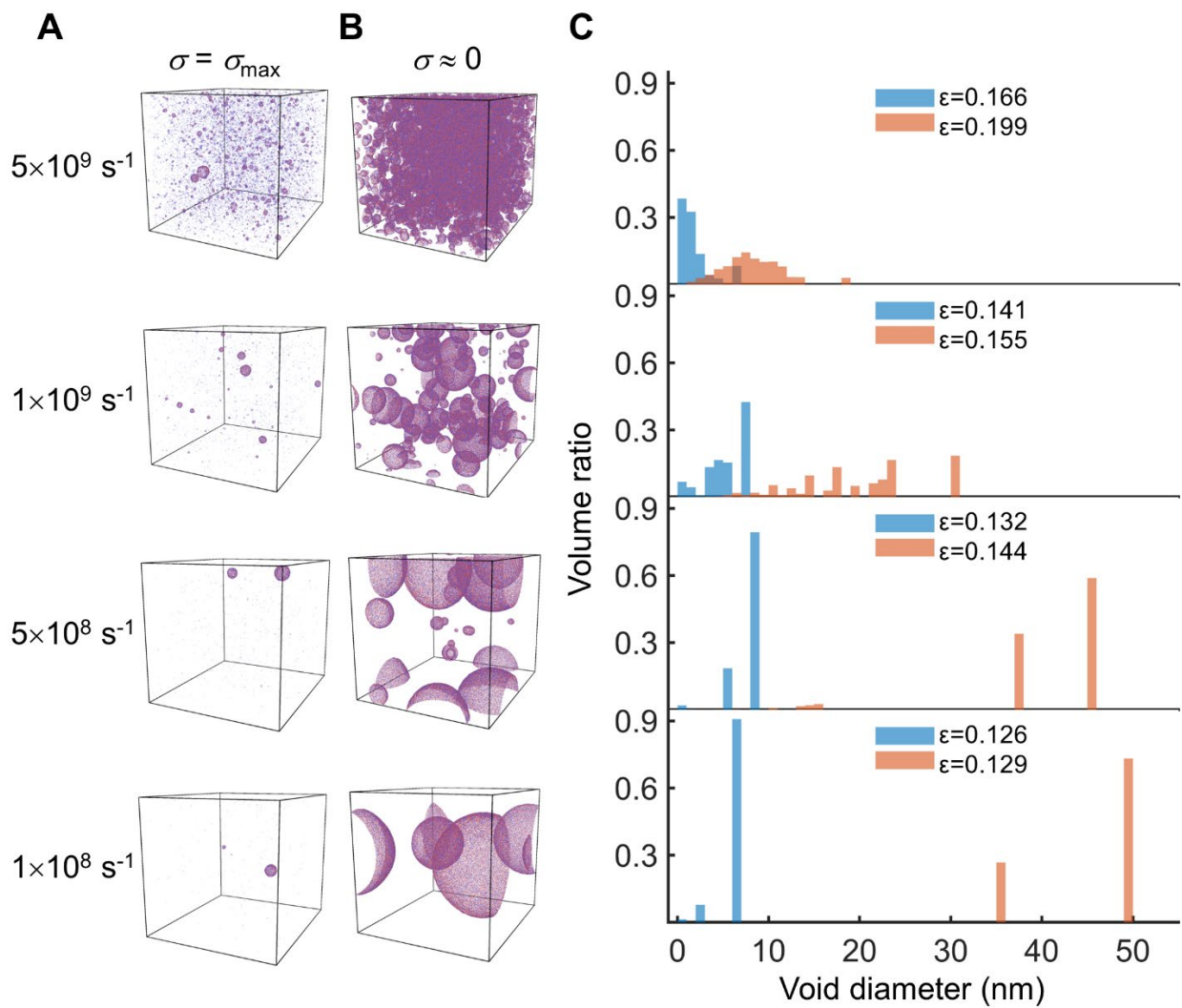

**Fig. S7.**
Rate-dependent void nucleation and growth revealed by molecular dynamics simulations. Void morphologies in $Cu_{50}Zr_{50}$ at the peak stress (**A**) and after material failure (i.e., when the stress decreases to zero) (**B**), at different strain rates. (**C**) Statistics of the void sizes at the peak stress and after material failure at different strain rates. The strains at two typical moments for each simulation are listed.

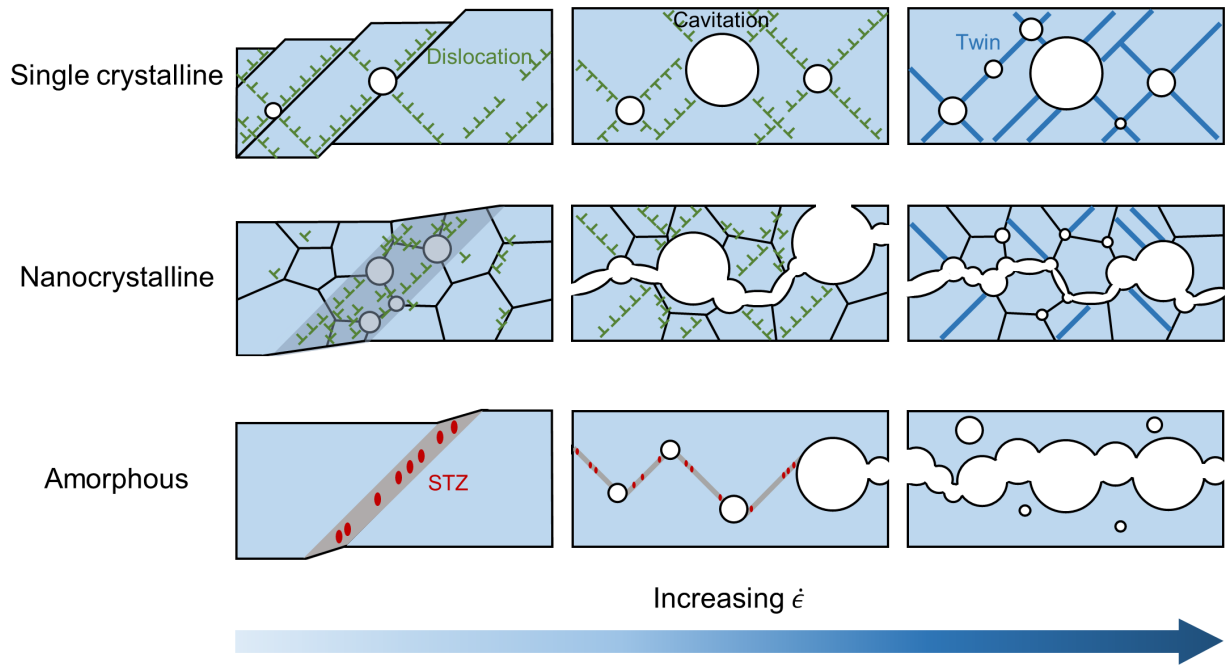

**Fig. S8.**
Failure mechanisms of crystalline and amorphous metals. Schematic diagrams of the representative failure mechanisms of metallic materials at various strain rates; arrows denote the tensile direction. Single crystals mainly fail due to slip bands mediated by multiple dislocations at quasi-static or low strain rates, cavitation induced by dislocations and emissions at high strain rates (*13*), and cavitation induced by intersections of twins at ultrahigh strain rates (*5*). Nanocrystalline metals fail due to similar microscopic mechanisms, except that the grain boundaries and triple junctions act as sinks and sources for dislocation activities (*5, 14*). In contrast, the main failure mechanism for amorphous metals is formation of shear bands at quasi-static or low strain rates (*30*). At high strain rates, amorphous metals fail as the result of concurrent shear band and cavitation mechanisms (*20*). Our study shows that amorphous metals fail mainly due to void growth and coalescence at ultrahigh strain rates (> $10^7$ s$^{-1}$).

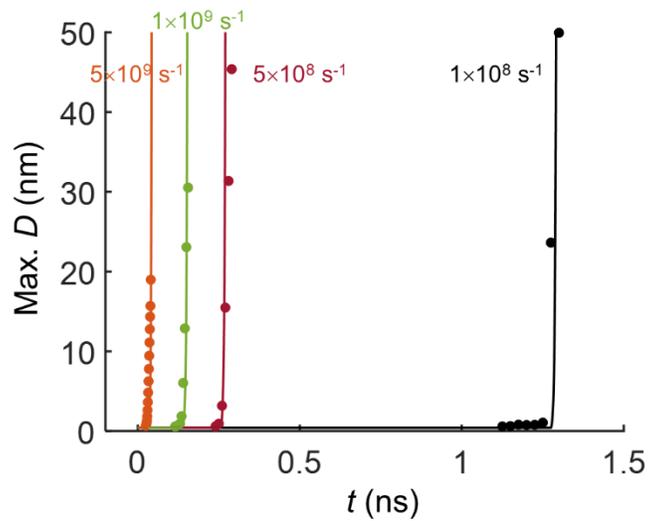

**Fig. S9.**
The temporal evolution of the maximum void diameter (symbols) at different strain rates during the MD simulation is in good agreement with the predictions of the kinetic model for void growth (lines).

| $\rho$ (kg/m³) | $c$ (m/s) | $v_{fsp}$ (m/s) | $\Delta v_{fs}$ (m/s) | $\Delta t$ (ns) | $\sigma_p$ (GPa) | $\sigma_s$ (GPa) | $\dot{\varepsilon}$ (1/s) | Refs. |
|---|---|---|---|---|---|---|---|---|
| 6000 | 5184 | 347 | 254 | 279 | 5.4 | 4.0 | $8.8\times10^4$ | Zhuang et al. |
| 6762 | 4768 | / | 162 | 77 | 3.6 | 2.6 | $2.2\times10^5$ | Escobedo et al. |
|  |  | / | 187 | 52 | 5.1 | 3.0 | $3.8\times10^5$ |  |
|  |  | / | 195 | 55 | 5.2 | 3.1 | $3.7\times10^5$ |  |
|  |  | / | 152 | 39 | 6.0 | 2.5 | $4.1\times10^5$ |  |
| 6760 | 4750 | 360 | 206 | 224, | 5.8 | 3.3 | $9.7\times10^4$ | Ding et al. |
|  |  | 480 | 260 | 159, | 7.7 | 4.2 | $1.7\times10^5$ |  |
|  |  | 520 | 310 | 175 | 8.4 | 5.0 | $1.9\times10^5$ |  |
| 6940 | 4740 | 453 363 | / | / | 7.4 | 4.1 | $7.0\times10^5$ | Tang et al. |
|  |  |  | / | / | 6.0 | 3.2 | $3.5\times10^5$ |  |
| 6301 | 4487 | 867 | 466 | 2.4 | 12.3 | 6.6 | $2.2\times10^7$ | Present study |
|  |  | 1340 | 578 | 2.7 | 18.9 | 8.2 | $2.4\times10^7$ |  |
|  |  | 992 | 399 | 2.9 | 14.0 | 5.6 | $1.5\times10^7$ |  |
|  |  | 1040 | 525 | 3.1 | 14.7 | 7.4 | $1.9\times10^7$ |  |
|  |  | 982 | 437 | 3.5 | 13.9 | 6.2 | $1.4\times10^7$ |  |
|  |  | 1695 | 638 | 3.2 | 24.0 | 9.0 | $2.2\times10^7$ |  |
|  |  | 1462 | 484 | 2.7 | 20.7 | 6.8 | $2.0\times10^7$ |  |
|  |  | 1753 | 525 | 2.1 | 24.8 | 7.4 | $2.8\times10^7$ |  |
|  |  | 1475 | 693 | 3.0 | 20.9 | 9.8 | $2.6\times10^7$ |  |

**Table S1.**

Summary of results from previous studies and laser-induced shock tests in the present study.

**References**

46. H. Shu *et al.*, Plastic behavior of aluminum in high strain rate regime. *Journal of Applied Physics* **116**, (2014).
47. P. Wen, B. Demaske, D. E. Spearot, S. R. Phillpot, Shock compression of CuxZr100−x metallic glasses from molecular dynamics simulations. *Journal of Materials Science* **53**, 5719-5732 (2017).
48. S. Plimpton, Fast parallel algorithms for short-range molecular dynamics. *Journal of Computational Physics* **117**, 1-19 (1995).
49. M. I. Mendelev, D. J. Sordelet, M. J. Kramer, Using atomistic computer simulations to analyze x-ray diffraction data from metallic glasses. *Journal of Applied Physics* **102**, (2007).
50. R. E. Ryltsev, B. A. Klumov, N. M. Chtchelkatchev, K. Y. Shunyaev, Nucleation instability in supercooled Cu-Zr-Al glass-forming liquids. *J Chem Phys* **149**, 164502 (2018).